\documentclass[aps,prl,preprint,showpacs,groupedaddress]{revtex4}
\usepackage {amssymb}
\usepackage {amsmath}
\usepackage {graphicx}
\usepackage {longtable}

\begin{document}
\title{Radial solitary waves in periodic variable stars}

\author{Fedor V.Prigara}
\affiliation{Institute of Microelectronics and Informatics,
Russian Academy of Sciences,\\ 21 Universitetskaya, Yaroslavl
150007, Russia} \email{fprigara@imras.net76.ru}

\date{\today}

\begin{abstract}

It is shown that the oscillations of brightness in classical
Cepheids and other 'pulsating' variable stars are caused rather by
the passing of radial solitary waves through the photosphere of a
star, than by pulsations of a star. Radial solitary waves also
determine the oscillations of brightness in some other types of
periodic variable stars ordinary not considered as pulsating
stars.

\end{abstract}


\maketitle

The periodic oscillations of brightness in classical Cepheids and
some other types of periodic variable stars are normally explained
by radial or non-radial pulsations caused by changes in absorption
of radiation in the layer of ionized helium beneath the star's
photosphere [1,2]. These pulsations models encounter, however,
essential difficulties. The maximum of brightness in classical
Cepheids does not coincide with the phase of maximum contraction
but, instead, coincides with the minimum value of radial velocity,
corresponding to the maximum rate of the photosphere expansion
[1]. Similarly, the minimum of brightness does not coincide with
the phase of maximum expansion, but, instead, corresponds to the
maximum rate of the photosphere's contraction. These properties of
classical Cepheids, as well as some other, can be, in a more
natural way, attributed to the passing through the star's
photosphere of radial solitary waves, causing the local heating or
cooling of the photosphere region and, also, the spatial
displacement of the photosphere. The crucial argument against the
pulsation models is that RR Lyrae stars show simultaneously both
the outward going and inward going radial solitary waves causing
the splitting of spectral lines in a star's spectrum near the
maximum of brightness. Note that the pulsation models become now
very complicated and nonlinear (see, e.g., [3]). Besides, the
pulsating stars seem to be only the minority of the whole class of
stars located in the instability strip of the Hertzsprung-Russell
diagram [1], the instability strip being determined by the
pulsation models of stars.

After passing the photosphere of a star, the outward going radial
solitary waves are reflected from the surface of three-dimensional
star's magnetosphere. The spherical form of this surface which is
required for the reflection of radial solitary waves indicates
that the magnetic field of a star is not simply a regular magnetic
field of the three-dimensional magnetic dipole. It is rather
formed by convection in the star's envelope above the photosphere.

In the case of classical Cepheids, the maximum of brightness
coincides with the maximum amplitude of outward going radial
solitary wave passing the photosphere region. It is clear that the
outward propagating radial solitary wave causes the local heating
of a plasma in the photosphere region. The successive radiative
cooling of a plasma leads to the smooth decrease of the
temperature in the photosphere region and, subsequently, to the
smooth decrease of brightness. The minimum of brightness coincides
with the maximum amplitude of inward propagating radial solitary
wave reflected from the outer surface of the star's magnetosphere.

Presumably, the outward propagating radial solitary wave in the
star's atmosphere is the ion-sound shock wave [4] produced by the
transformation of acoustic type wave on the boundary between the
dense core of a star and its more rarified envelope.

In the case of RR Lyrae stars, the outward going and inward going
radial solitary waves exist simultaneously. Both of them are seen
not long before the maximum of brightness, causing the splitting
of spectral lines in the star's spectrum [1]. From the time
dependence of the radial velocities, it is clear that the period
of the radial velocity oscillations in RR Lyrae stars is twice the
(minimum) period of the brightness oscillations [1]. It means that
the true light curve of a RR Lyrae star has two maximums of
brightness, similar to the light curves of RV Tauri stars. In
general, there is a sequence of types of periodic variable stars,
along which the light curves are smoothly transformed each in
other: $\delta$ Cephei - RR Lyrae - RV Tauri - $\beta$ Lyrae -
$\beta$ Persei. Note that RR Lyrae stars often have two periods of
the brightness oscillations, similarly to RV Tauri stars.

The last two classes of periodic variable stars in the above
sequence are ordinary considered as non-pulsating. However, many
phenomena seen in these types of variable stars, such as the
change of period and the presence of emission lines, are common
with the classes of physical variable stars and support the
suggestion that the oscillations of brightness in $\beta$ Lyrae
and $\beta$ Persei stars are also caused by the outward going and
inward going radial solitary waves passing through the photosphere
region of a star.

In the $\beta$ Lyrae system, the second component shows no
absorption lines at any phase, including the phases of suggested
eclipses at the minima of brightness, despite its suggested large
luminosity [5]. Contrary, the B8 spectrum of the principal
component of this system is observed at every phase. The minima of
brightness in the $\beta$ Lyrae system correspond to approximately
zero values of the radial velocity shown by B8 absorption lines,
and each of the two maxima correspond to the minimum (the first)
or the maximum (the second) of the velocity curve [5]. It is clear
that the outward going radial solitary wave is responsible for the
first maximum of brightness, corresponding to the maximum rate of
expansion of the star's photosphere, while the inward going radial
solitary wave is responsible for the second one, corresponding to
the maximum rate of the photosphere's contraction. The principal
minimum of brightness coincides with the reflection of the radial
solitary wave from the central region of the star, and the
secondary minimum of brightness coincides with the reflection of
the radial solitary wave from the surface of the three-dimensional
magnetosphere of the star. The $\beta$ Lyrae system invokes also
an accretion disk producing in particular CIV emission lines and a
convective envelope producing P Cyg type lines [1].

Earlier, it was shown that pulses of radio emission from a radio
pulsar are triggered by radiation-induced radial solitary waves
propagating in hot plasmas of the accretion disk surrounding a
pulsar [6]. Periodic variable stars also have accretion disks,
indicative of which is the presence of emission lines in their
spectra. The outward going radial solitary wave propagating in the
star's three-dimensional envelope can produce the outward going
radial solitary wave in the accretion disk surrounding a star. The
last wave causes the occurrence of emission lines in the star's
spectrum corresponding to some phases of the star's light curve.

In the case of UV Ceti stars undergoing flares of optical emission
[1], the ultraviolet/optical bump is seen in their spectra during
the flare. This bump is characteristic for an accretion disk and
is normally present in quasars [7]. The radio emission from UV
Ceti stars following the optical flare can be interpreted as the
radio emission triggered by the outward going radial solitary wave
in an accretion disk surrounding a star [6].

To summerise, outward- going and inward- going radial solitary
waves in the star's envelope are responsible for the oscillations
of brightness, the splitting of spectral lines, and the occurrence
of emission lines in periodic variable stars of various types,
both 'pulsating' and 'non-pulsating'.

\begin{center}
---------------------------------------------------------------
\end{center}

[1] C.Hoffmeister, G.Richter, and W.Wenzel, \textit{Variable
Stars} (Springer- Verlag, New York, 1988).

[2] A.G.Masevich and A.V.Tutukov, \textit{The Evolution of Stars:
Theory and Observations} (Nauka, Moscow, 1988).

[3] M.Marconi, T.Nordgren, G.Bono, G.Schnider, and F.Caputo,
Astrophys. J. \textbf{623}, L133 (2005).

[4] F.F.Chen, \textit{Introduction to Plasma Physics and
Controlled Fusion, Vol.1: Plasma Physics} (Plenum Press, New York,
1984).

[5] O.Struve, Publ. Astron. Soc. Pac. \textbf{70}, 5 (1958).

[6] F.V.Prigara, in \textit{Gravitation, Cosmology and
Relativistic Astrophysics}, edited by Yu.V.Alexandrov et al.
(Kharkov National University, Kharkov, 2004), pp. 77-81.

[7] H.Falcke, E.Koerding, and S.Markoff, Astron. Astrophys.
\textbf{414}, 895 (2004).

\end{document}